\UseRawInputEncoding
\documentclass[namedreferences]{solarphysics}
\usepackage[hyperref,optionalrh,solaromanenum]{spr-sola-addons} 
\usepackage{graphicx}                    
\usepackage{amssymb}                    
\usepackage{color}                       
\usepackage{breakurl}
\usepackage{array}                   

\begin{document}


\begin{opening}

\title{Separating aa-index into Solar and Hale Cycle Related Components Using Principal Component Analysis}

%
\author[addressref={aff1},corref,email={jojuta@gmail.com}]{\inits{J.T.}\fnm{Jouni}~\lnm{Takalo}}

\institute{$^{1}$ Space Physics and Astronomy Research Unit, University of Oulu, POB 3000, FIN-90014, Oulu, Finland}
%
\runningauthor{J.J. Takalo}
\runningtitle{Decomposing aa-index using PCA}



\begin{abstract}

We decompose the monthly aa-index of cycles 10\,--\,23 using principal component analysis (PCA). We show that the first component (PC1) is related to solar cycle, and accounts for 41.5\, \%  of the variance of the data. The second component (PC2) is related to 22-year Hale cycle, and explains 23.6\,\% of the variance of the data. The PC1 time series of aa cycles 10\,--\,23 has only one peak in its power spectrum at the period 10.95 years, which is the average solar cycle period for the interval SC10\,--\,SC23. The PC2 time series of the same cycles has a clear peak at period 21.90 (Hale cycle) and a smaller peak at 3/4 of that period. 
We also study the principal component of sunspot numbers (SSN) for cycles 10\,--\,23, and compare mutual behavior of the PC2 components of aa-index and SSN PCA analyses. We note that they are in the same phase in all other cycles than Solar Cycles 15 and 20. The aa cycle 20 also differs from other even aa cycles in its shape, especially in anomalously high peaks during its descending phase.   
Even though there is a coherence in the PC2 time series phases of aa-index and sunspot number, this effect is too small to cause all of the difference of the shape of even and odd aa cycles. We estimate that 30\,\% of the shape of the PC2 component of aa-index is due to sunspot number and the rest to other recurrent events in Sun/solar-wind. The first maximum of aa-data (typical to odd cycles), during sunspot maximum, has been shown to be related to coronal mass ejections (CME), while the second maximum (typical to even cycles) in the declining phase of cycle, is probably related to high-speed streams (HSS). The last events increase the activity level such that the minimum between even-odd cycle-pair is always higher than the minimum between succeeding odd-even cycle pair. 

\end{abstract}


\end{opening}

\section{Introduction}

\begin{figure}
\centering
	\includegraphics[width=0.8\textwidth]{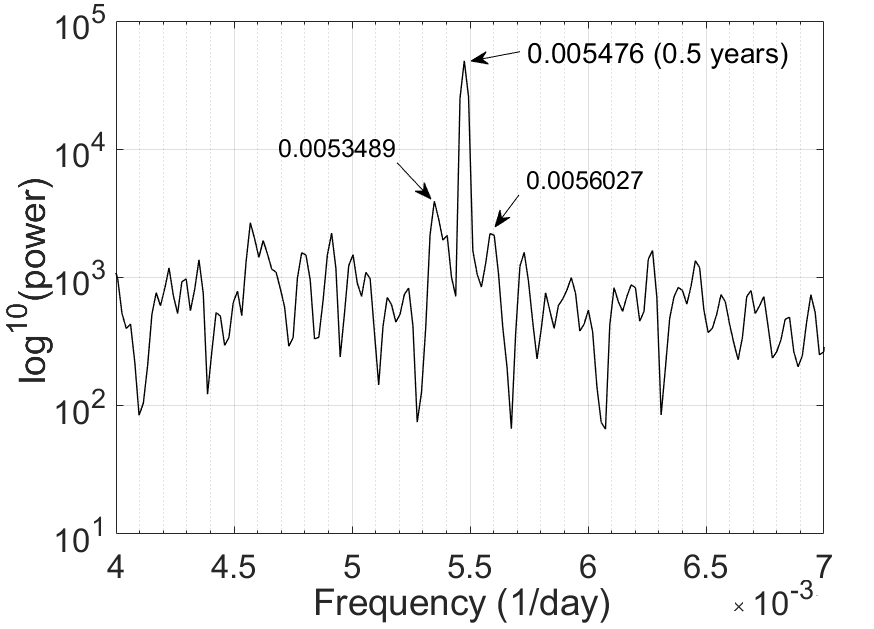}
		\caption{The amplitude modulation of the semiannual peak with the 22-year period in the daily aa-index of 1868\,--\,2018.}
		\label{fig: semiannual modulation}
\end{figure}

The 22-year periodicity or double sunspot cycle in the geomagnetic activity has been studied since \cite{Chernosky_1966} noticed that
international geomagnetic index Ci revealed characteristically different patterns in even- and odd-numbered solar cycles. This cycle is
found in the recurrence of the 27-day geomagnetic activity index Ap and ionospheric F2 layer variability \citep{Apostolov_2004}. \cite{Shnirman_2009} applied wave packet technique to geomagnetic aa-index and found 22-year variation in the interval 25-31 days centered at 27-day solar rotation period.
More commonly, the 22-year modulation is observed in the semiannual variation in the geomagnetic indices. This modulation has been attributed to three various explanations: axial mechanism \citep{Cortie_1912, Bohlin_1977}, equinoctial hypothesis \citep{Bartels_1932, McIntosh_1959, Svalgaard_1977}, and Russell-McPherron effect \citep{Russell_1973} (see also reviews \citep{Cliver_2000, Cliver_2002}).
The main reason for this 22-year variation is not the scope of this article. In Figure \ref{fig: semiannual modulation} we, however, show the amplitude modulation of the semiannual peak with the 22-year period in the daily aa-index of 1868-2018. The highest peak corresponds to the half-year frequency and the other two peaks in the both sides of the main peak are side peaks due to the modulation. According to the amplitude modulation, the frequency difference of the side peaks is twice the modulation frequency \citep{Takalo_1995}. That is why we can calculate the modulation period as 2/(0.0056027-0.0053489)=7880 days=21.6 years.\newline
There have been some studies about decomposing aa-index to separate components \citep{Feynman_1982, Echer_2004, Hathaway_2006, Du_2011_2}. These studies use mostly annually averaged aa-index. \cite{Feynman_1982} analyzed the relationship between the annual aa and sunspot number R series from 1869 to 1975 and found that the aa-values are all above a baseline ($aaR$) that is linearly related to R . Then, she decomposed the aa-index into two equally strong periodic components: $aaR$ and the remainder $aaI = aa-aaR$. The $aaR$ component is associated with the transient phenomena and follows the sunspot cycle, while the $aaI$ (interplanetary) component is associated with the recurrent phenomena and is almost 180 degrees out of phase with the sunspot cycle. \cite{Hathaway_2006} presented quite similar decomposition to that of \cite{Feynman_1982}. One component is proportional to, and in phase with, the sunspot number and another, interplanetary, component $aa_{I}$, which is out of phase with the sunspot cycle. This second component peaks some years before solar minimum, and is one of the most reliable indicators for the amplitude of the following solar maximum. \cite{Echer_2004} stated that aa-index has a clear double-peak structure and decomposed the aa to sunspot number [$R_{z}$] and fast solar-wind (SW$>$500 $kms^{-1}$) related terms. They found that 71\,\% of the variability in aa-index can be explained by a linear dependency on these terms.
\cite{Du_2011_2} presented two different ways to decompose the aa-index into components. The first model was similar to the aa model by \cite{Feynman_1982}, but the components are (almost) perpendicular (90 degree phase shift) to each other. The second model is based on the logarithmic relationship between annual values of aa-index and $R_{z}$. According to this model all aa-values are between the lines $aa_{t}= e^{2.44}R_{z}^{2/7}$ (top-line) and $aa_{b}= e^{1.36}R_{z}^{2/7}$ (baseline), and aa-index can be decomposed to independent terms using these lines (see more profound presentation in \citep{Du_2011_2}).

In this study we present a new way to decompose the aa-index into components using principal component analysis (PCA). We extract the main three principal components from seven month trapezoidal-smoothed aa data. The first component is due to the 10.95 year solar cycle, and the second component is related to 21.90 year Hale cycle. According to the definition of PCA, these components are mutually orthogonal. The third (and upper) PCs show shorter periods and are related to features of some specific cycles. This paper is organized as follows. Section 2 presents the data and methods used in this study. In Section 3 we present the results of PC analysis for aa-index and sunspot number index of the Solar Cyles 10\,--\,23 and discuss their connection in Section 4. We give our conclusion in Section 5.

\section{Data and methods}

\subsection{aa-index}

\cite{Mayaud_1972} presented geomagnetic activity aa-index, which is based on the K indices of two antipodal stations, one in Australia and another in southern England, which started measurements 1868.  For the northern hemisphere the sites have been Greenwich (for 1868\,--\,1925), Abinger (1926\,--\,1956) and Hartland (1957\,--\,present), and for the southern hemisphere they are Melbourne (1868\,--\,1918), Toolangi (1919\,--\,1979) and Canberra (1980\,--\,present). Later this index has been extended for two solar cycles between 1844\,--\,1868 using measurements made in Helsinki \citep{Nevanlinna_1993, Lockwood_2013}. There exists a provisional aa-index also until the end of year 2020 (http://isgi.unistra.fr), but the main analysis is made for the aa cycles 10\,--\,23 (from August, 1856 to December, 2009), because the aa cycle 24 is still incomplete, and for a complete Hale cycle analysis we need also the next odd cycle 25.
 
\begin{figure}
\centering
	\includegraphics[width=1.0\textwidth]{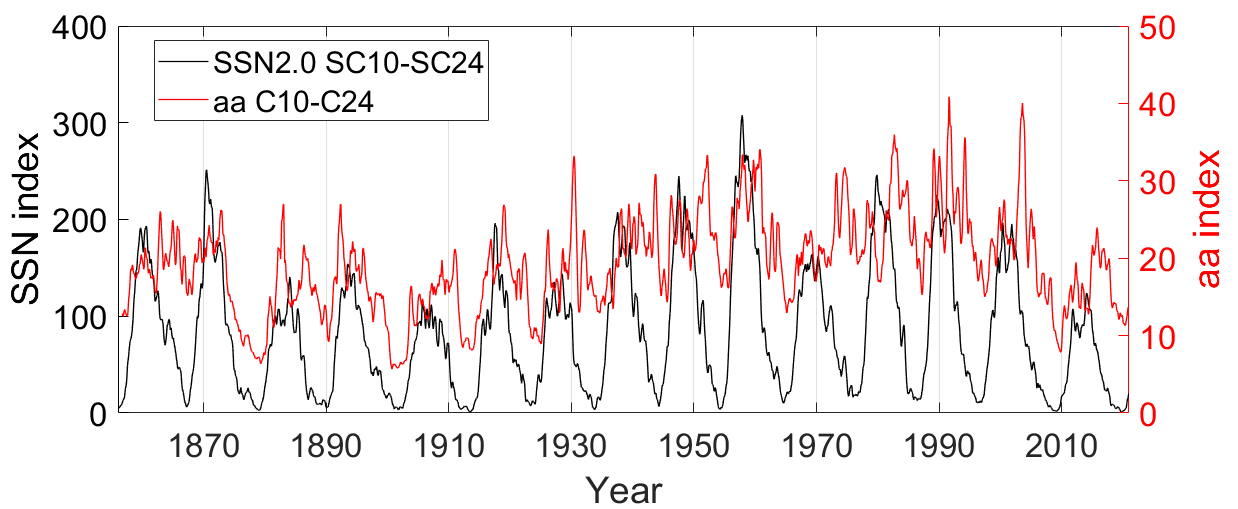}
		\caption{Comparison of aa- and SSN2-indices. Black line shows the SSN2-index (y-axis left), and red line shows the aa-index (y-axis right). Notice, that aa-index maxima (minima) lack the maxima (minima) of the sunspot number index.}
		\label{fig:SSN2_aa}
\end{figure}

The aa-index is, however, quite noisy and we have made seven-month trapezoidal smoothing for the index before analysis. This also removes the semiannual variation in the aa-index. Trapezoidal smoothing is a moving average smoothing such that the end points of the window have only half of the weight from the other points. Also the cycles for aa-index lag somewhat those for SSN index. This is shown in Fig.\ref{fig:SSN2_aa}, where seven-month smoothed sunspot number (SSN2.0, http://www.sidc.be/silso/) and aa-index are plotted together. Note that the difference between aa cycles and solar cycles is largest in the middle of the period SC10\,--\,SC23, i.e. for the cycles 16\,--\,19 corresponding the years 1923\,--\,1955. The correlation analysis, using seven-month smoothed indices, showed that the mean lag between aa-index and SSN-index for the cycles C10\,--\,C23 is about 10\,--\,11 months. In order to determine the minima of the aa cycles 10\,--\,23, we used 13-month trapezoidal smoothed aa time series. (Note that 13-month smoothing is used only for determining the aa-index minima, otherwise we use seven-month smoothing for both aa-index and SSN-index.) The lags of these minima compared to SSN mimima are shown in the last column of Table 1. Note that aa time series starts eight months later than SSN time series as shown earlier in the Table1. It is also clear that the aa cycle minima lag solar cycle minima most in the middle of 20th century, i.e., for the cycles 16\,--\,19 \citep{Kane_2007}.

\begin{table}

\caption{Sunspot cycle lengths and dates of starting minima used in the calculation of the analysis. The last column shows the lag of corresponding aa cycle minima.}
\begin{tabular}{ |c|c|c|c|}
\hline
   Sunspot cycle    &Year and month  &Cycle length  &Lag of corresponding \\
	    number        &of starting min   &(years)      & aa cycle minimum  \\
\hline

10       &1855 December  & 11.2  & +8\\
11       &1867 March  & 11.8  & +8\\
12       &1878 December  & 10.6  & +5\\
13       &1889 August  & 12.1  & +14\\
14       &1901 September  & 11.8 & +3 \\
15       &1913 July  & 10.1  & +4\\
16       &1923 August & 10.1 & +17 \\
17       &1933 September  & 10.4  & +11\\
18       &1944 February  & 10.2  & +16\\
19       &1954 April  & 10.5  & +15\\
20       &1964 October  & 11.7  & +9\\
21       &1976 June & 10.2  & +9\\
22       &1986 September  & 10.1  & +6\\
23       &1996 October  & 12.2  & +12\\
24       &2008 December  &    & +12   \\

\hline
\end{tabular}

\end{table}

\subsection{Solar sunspot index}

Rudolf Wolf collected the Z\"urich series of sunspot numbers (SSN), which started in 1749. The first complete sunspot cycle included in SSN started in March 1755. Wolf started the numbering of sunspot cycles from this cycle and this numbering is still in use. The initial sunspot number series (SSN1) was reconstructed at the Z\"urich Observatory until 1980, and at the Royal Observatory of Belgium since 1981. Following the change of the reconstruction method in 1981, the current version of the SSN series is called the international sunspot number (ISN). Recently the ISN series was modified to a version 2.0 (SSN2) that is supposed to present a preliminary correction of the known inhomogeneities in the SSN1 series \citep{Clette_2014}. We have used the version SSN 2.0 in this study, and call it later just SSN.

\subsection{Principal component analysis method}

Principal component analysis is a useful tool in many fields of science, chemometrics \citep{Bro_2014}, data compression \citep{Kumar_2008} and information extraction \citep{Hannachi_2007}. PCA finds combinations of variables, that describe major trends in the data. PCA has earlier been applied, e.g., to studies of the geomagnetic field \citep{Bhattacharyya_2015}, geomagnetic activity \citep{Holappa_2014_2, Holappa_2014_1}, ionosphere \citep{Lin_2012}, and the solar background magnetic field \citep{Zharkova_2012, Zharkova_2016}. In this article we estimate that the average length of the cycle is 131 months, and use it as a representative solar cycle model. We also use this same length (131 months) as a common length for aa-index cycles. To this end, we first resample the monthly sunspot values (aa values) so that all cycles have  the same length of 131 time steps (months). This effectively elongates or abridges the cycles to the same length. Before applying the PCA method to the resampled sunspot cycles (aa cycles) we standardize each individual cycle to have zero mean and unit standard deviation. This guarantees that all cycles will have the same weight in the study of their common shape. Then after applying the PCA method to these resampled and standardized cycles, we revert the cycle lengths and amplitudes to their original values (see also \citep{Takalo_2018}).

Before we apply PCA to aa data we standardize time series each individual cycle of aa-index to have zero mean and unit standard deviation. Standardized data are then collected into the columns of the matrix $X$, which can be decomposed as \citep{Hannachi_2007, Holappa_2014_1, Takalo_2018}

\begin{equation}
	X = U\:D\;V^{T}  \     ,
\end{equation}

where $U$ and $V$ are orthogonal matrices, $V^{T}$ a transpose of matrix $V$, and $D$ a diagonal matrix 
	$D= diag\left(\lambda_{1},\lambda_{2},...,\lambda_{n}\right)$
with $\lambda_{i}$ the $i^{th}$ singular value of matrix $X$. The principal component are obtained as the the column vectors of

\begin{equation}
P  = U\!D.
\end{equation}
	
The column vectors of the matrix $V$ are called empirical orthogonal functions (EOF) and they represent the weights of each principal component in the decomposition of the original normalized data of each cycle $X_{i}$, which can be approximated as

\begin{equation}
	X_{i} = \sum^{N}_{j=1} \:P_{ij}\:V_{ij} \   ,
\end{equation}

where j denotes the $j^{th}$ principal component (PC). The explained variance of each PC is proportional to square of the corresponding singular value
$\lambda_{i}$. Hence the $i^{th}$
PC explains a percentage
\begin{equation}
\frac{\lambda^{2}_{i}}{\sum^{n}_{k=1}\!\lambda^{2}_{k}} \cdot\:100\%
\end{equation}
of the variance in the data.

\section{Results}

\subsection{PCA analysis of aa-index}

\begin{figure}
	\centering
	\includegraphics[width=0.75\textwidth]{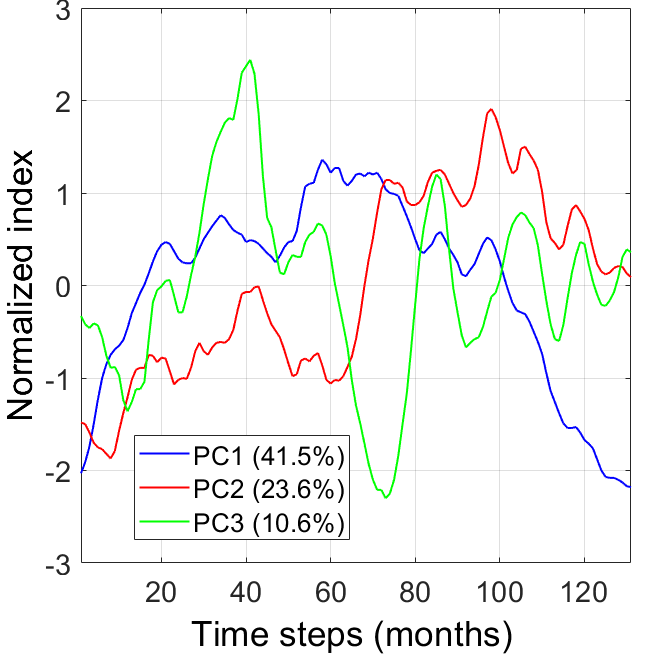}
		\caption{The three first PCs of the aa-index for cycles 10\,--\,23.}
		\label{fig:aa_PCs}
\end{figure}

Using modified (solar related) cycles for aa-index C10-C23, we made PCA analysis by equalizing the aa cycles to 131 time steps (months) to get the three main principal components shown in Fig. \ref{fig:aa_PCs}. The first, second and third PC explain 41.5\,\% , 23.6\,\% and 10.6\,\% of the total variation of the data, respectively. Hence the three main PCs account together for 75.7\,\% of the smoothed aa-index. Although PC1 accounts only for 41.5\,\% of the variance of aa-index, the PC1 has correlation coefficient 0.998 (p < $10^{-100}$) with the mean aa cycle (C10\,--\,C23). Note, however, that PC2 explains almost a quarter of the total variance in the aa-index. 

\begin{figure}
	\centering
	\includegraphics[width=0.9\textwidth]{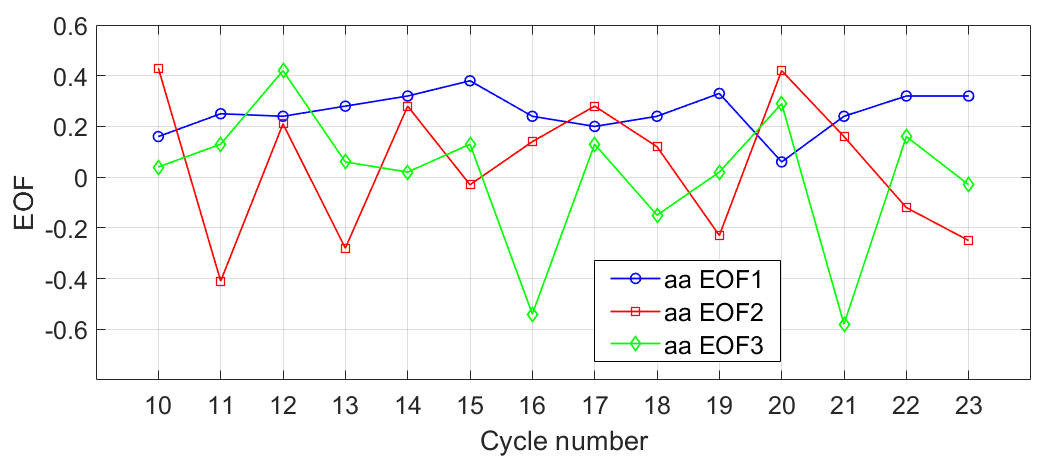}
		\caption{The three first EOFs of the aa-index for cycles 10\,--\,23.}
		\label{fig:aa_EOFs}
\end{figure}

Figure \ref{fig:aa_EOFs} shows the three EOFs of the aa cycles. Note especially the almost sawtooth like shape of the EOF2. All even numbered EOF2s are positive, except for cycle 22, and all the odd numbered EOF2s are negative, except for cycle 17, and cycle 21. Looking the shape of aa PC2, we note that a positive EOF2 means positive phase in the second half of the cycle, i.e., in the declining part of the the average, and a negative EOF2 positive phase in the ascending part of the cycle. 

\begin{figure}
	\centering
	\includegraphics[width=0.95\textwidth]{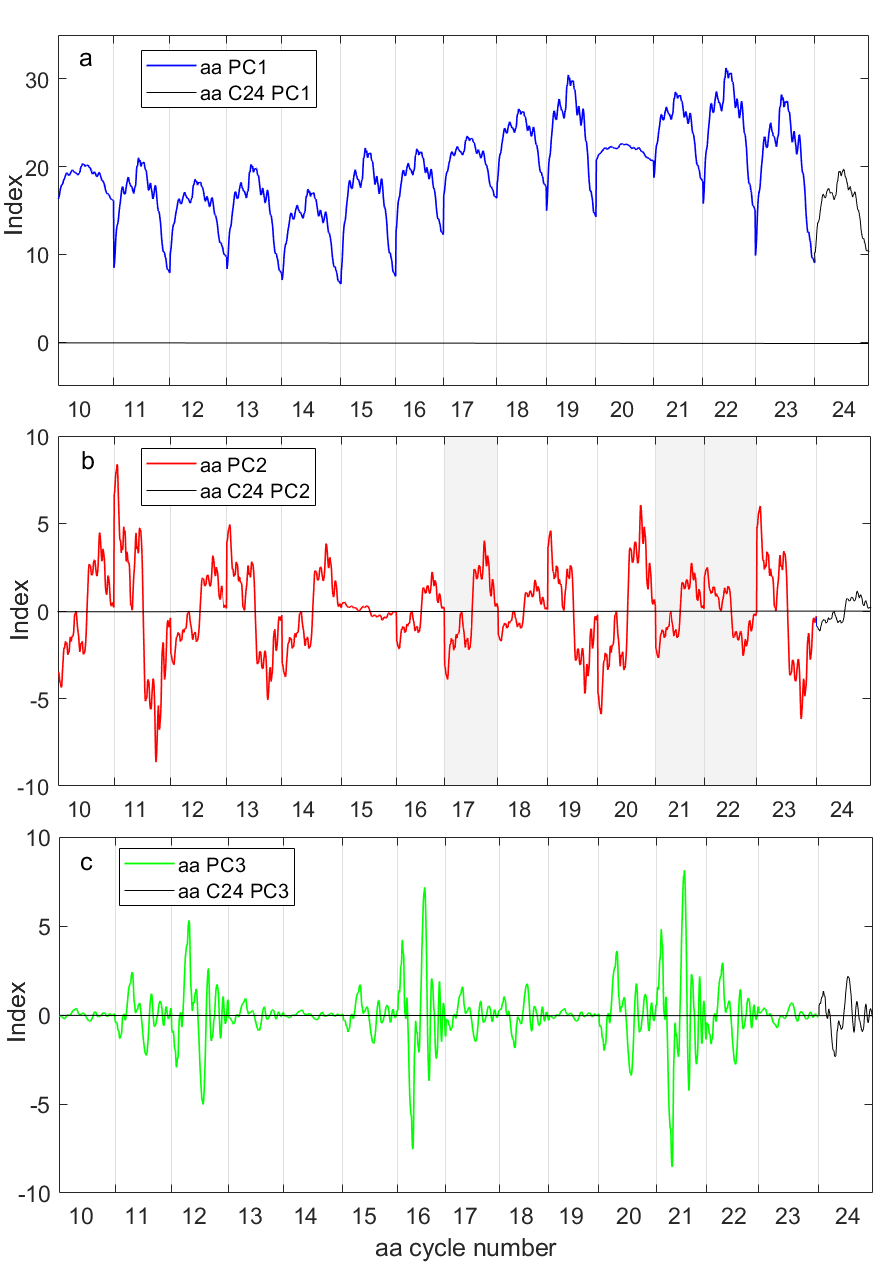}
		\caption{a PC1 b) PC2, and C) PC3 time series of the PCA of aa-index for cycles 10\,--\,23. Note that cycle 24 PCs are calculated from PCA of cycles 10\,--\,24.}
		\label{fig:aa_PC_timeseries}
\end{figure}

Reverting the PC1, PC2 and PC3 cycles back to their original length and concatenating them we get first, second and third principal component time series for the aa cycles of C10\,--\,C23 shown in Fig. \ref{fig:aa_PC_timeseries}a, b, and c, respectively. It is evident, that the PC1 is dominated with the sunspot cycle related period. Interestingly, the size of PC1 is almost equal up to cycle 19, although the accumulation of the activity level raises background and the total height of the index. Cycle 20 seems to have smallest PC1 component (although it height is similar to heights of the PC1s of cycle 16 and 17 due to the background baseline) and after that PC1 gradually increases for cycles 21\,--\,23. Note that at the end of Fig. \ref{fig:aa_PC_timeseries} time series we show the cycle 24 components as calculated from PCA of cycle C10\,--\,C24. It is evident that the shape of PCs of cycle 24 do not differ much from the other PCs, although with quite low amplitude.

In contrast with PC1, the aa PC2 shows clear Hale cycle period (see Fig. \ref{fig:aa_PC_timeseries}b). Note that the consecutive even-odd cycle pairs show similar structure for three Hale cycles (Hale cycle is traditionally referred as even-odd cycle pair \citep{Gnevyshev_1948,Wilson_1988, Makarov_1994, Cliver_1996}), i.e. for pairs 10\,--\,11, 12\,--\,13, and 14\,--\,15, although cycle 15 has quite small PC2 component. The cycles 16\,--\,17 have, however, mutually same structure , i.e. they are in the same phase, which could be seen also from the EOF2 of the odd cycle 17 (the first odd cycle with positive EOF2). The cycle-pair 18\,--\,19 is again similar to earlier Hale cycles, but the cycles 20\,--\,21 are again in the same phase (C21 is another odd cycle with positive EOF2). The PC2 of the cycle 22 is the only even cycle with positive phase in the first half and negative phase in the second half of the cycle., i.e. cycle 22 is the only even cycle with negative EOF2. Cycle 23 is a common odd-cycle with negative EOF2. It seems that cycle 24 is again common even aa cycle (PC2 is similar to the common even cycle PC2, although with quite low amplitude), but as stated earlier we need next odd cycle 25 to have a complete even-odd Hale cycle for our analysis.

\begin{figure}
	\centering
	\includegraphics[width=1.0\textwidth]{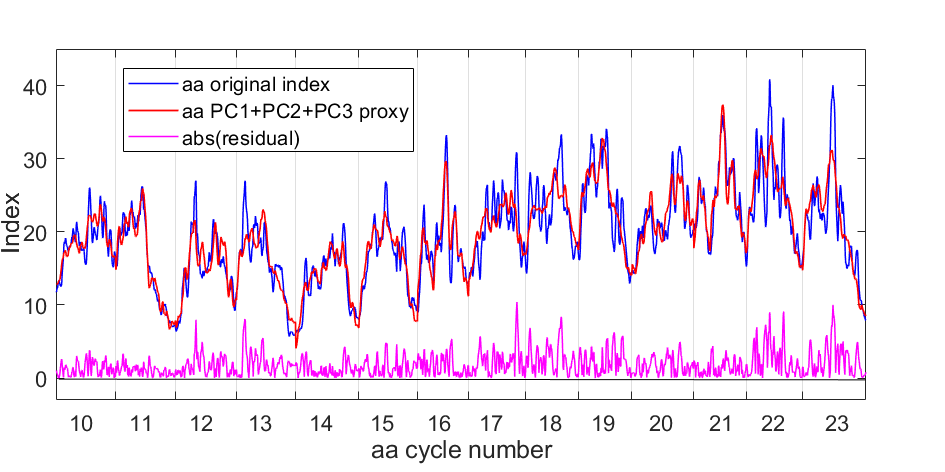}
		\caption{Original monthly aa time series (blue), its PC1+PC2+PC3 proxy time series (red), and absolute value of their difference (residual) time series (magenta) for cycles 10\,--\,23.}
		\label{fig:aa_and_PC_proxy}
\end{figure}

The PC3 time series of Fig. \ref{fig:aa_PC_timeseries}c shows higher order periods. Note, that there are four cycles (C10, C14, C19, and C23) with very small PC3 component. The higher PC-components usually describe some special features of only a few cycles. Figure \ref{fig:aa_and_PC_proxy} shows the original aa-index time series, PC1+PC2+PC3 proxy time series, and the residual time series. Notice that some high peaks are cut out from the PC1+PC2+PC3 proxy, because they exist only in some individual cycles.

\begin{figure}
	\centering
	\includegraphics[width=1.0\textwidth]{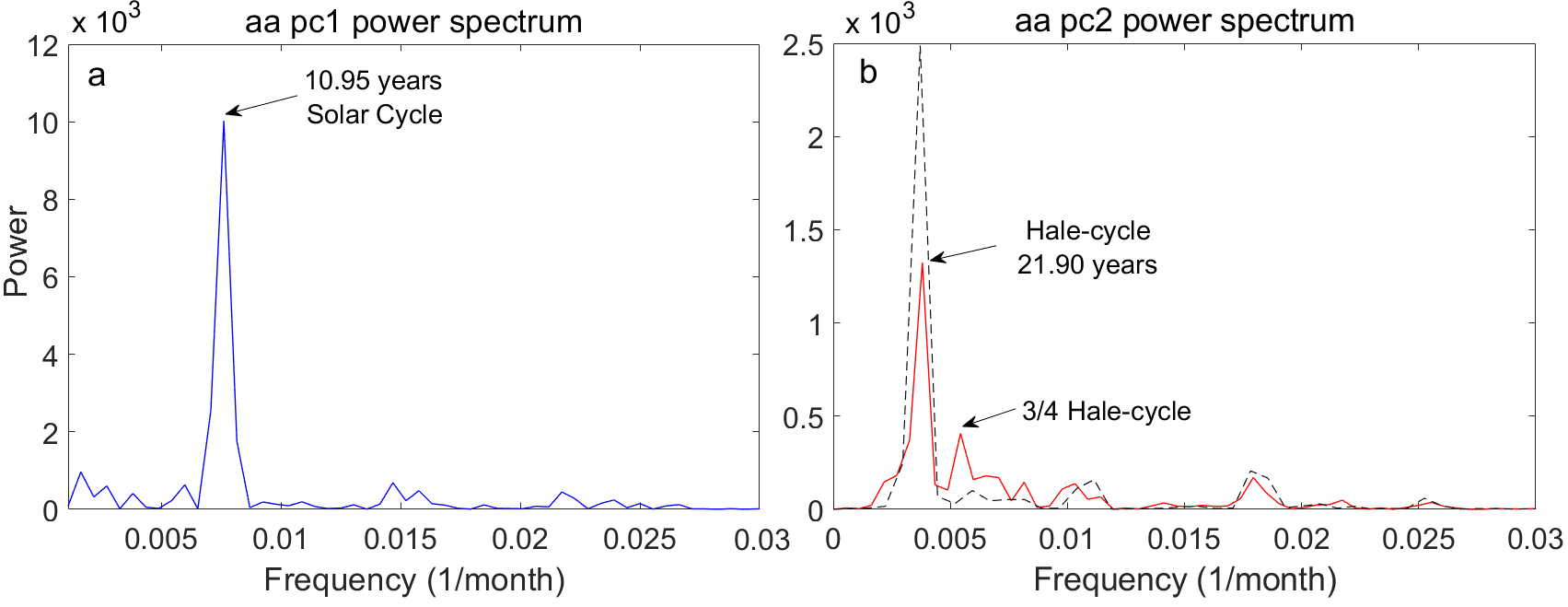}
		\caption{The power spectra of the a) PC1 and b) PC2 time series of Figure \ref{fig:aa_PC_timeseries}.}
		\label{fig:PC1_PC2_power_spectra}
\end{figure}

Figure \ref{fig:PC1_PC2_power_spectra}a and b show the power spectra of PC1 and PC2 time series of Fig. \ref{fig:aa_PC_timeseries}a, and b, respectively. It is clear that PCA effectively decomposes the solar cycle and Hale cycle (even-odd cycle) related periods from the aa-index. The solar cycle period in aa-index is 10.95 years and the Hale cycle 21.90 years. In the Fig. \ref{fig:PC1_PC2_power_spectra}b, there is another smaller peak at period of three quarters of the Hale cycle. We believe that this peak is due to the cycles where the phase of the common even-odd sequence changes (the cycles with ``wrong" phase are marked by light gray in the Fig. \ref{fig:aa_PC_timeseries}b). If we remove the incomplete Hale cycles 16\,--\,17 and 21\,--\,22 such that succeeding cycle-pairs are all in opposite phases (note that now the last pair is 20\,--\,23) we get only a higher peak at period 21.90 (shown with a dashed black line in the Fig. \ref{fig:PC1_PC2_power_spectra}b.

\subsection{Detrended aa-index}

\begin{figure}
	\centering
	\includegraphics[width=0.95\textwidth]{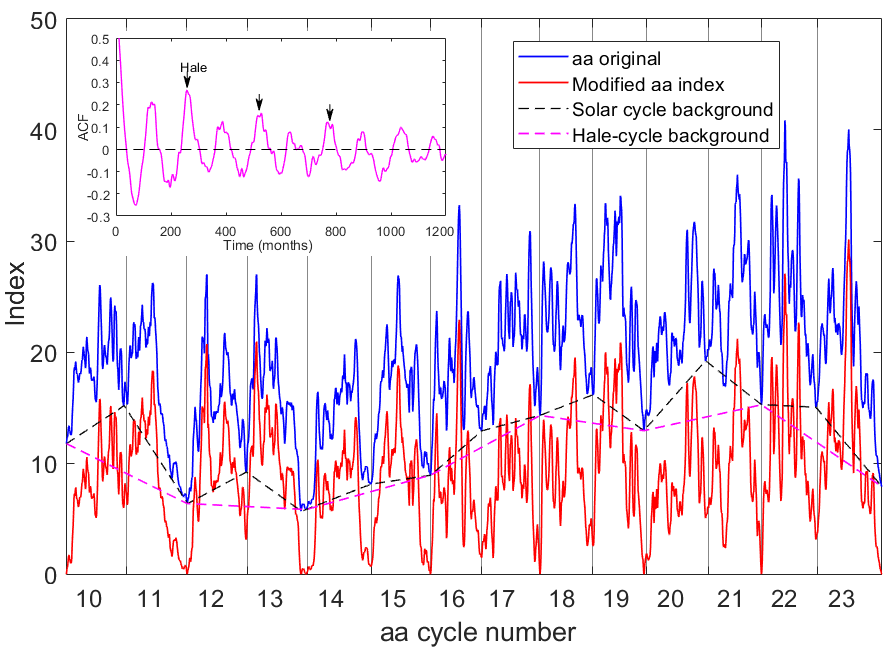}
		\caption{The original aa-index (blue), and the detrended aa-index (red) of the cycles 10\,--\,23. The inset shows the ACF of detrended aa-index.}
		\label{fig:modified_aa_and_its_acf} 
\end{figure}

Figure \ref{fig:modified_aa_and_its_acf} shows the original smoothed aa-index (red) and two background envelopes: Solar cycle background, which goes through all minima (dashed black poly-line), and Hale cycle background, which goes through the minima between odd and even aa cycles (magenta dashed poly-line). When subtracting the Hale cycle background envelope from the aa-index (blue) we get a detrended aa-index (red line). It is evident that there exist succeeding even-odd cycle pairs. The minima between the odd-even pairs are always higher than than the line drawn through the minima between the even-odd pairs. Furthermore, when long-term trend is removed from the aa-index index, the autocorrelation function (ACF) shows better the mutual intensity of the succeeding cycles. The inset of the figure \ref{fig:modified_aa_and_its_acf} shows ACF of the modified aa-index (magenta line, when Hale cycle background is subtracted). Note that every second peak is higher in the ACF showing Hale cycle in the aa-index.

\subsection{PCA of SSN cycles 10\,--\,23}

\begin{figure}
	\centering
	\includegraphics[width=0.75\textwidth]{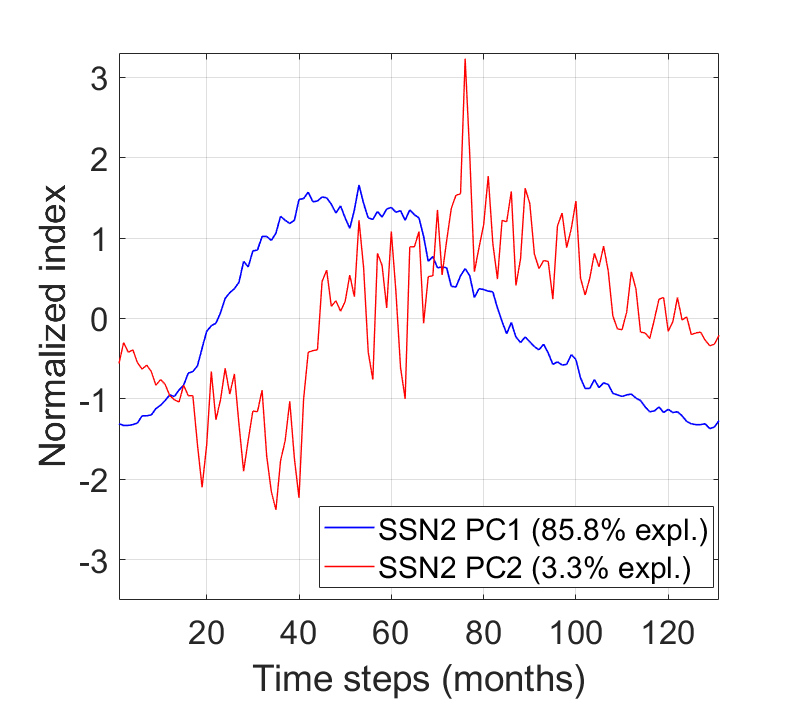}
		\caption{The PC1 and PC2 of SSN for the Solar Cycles 10\,--\,23.}
		\label{fig:R_PCs}
\end{figure}

\begin{figure}
	\centering
	\includegraphics[width=0.95\textwidth]{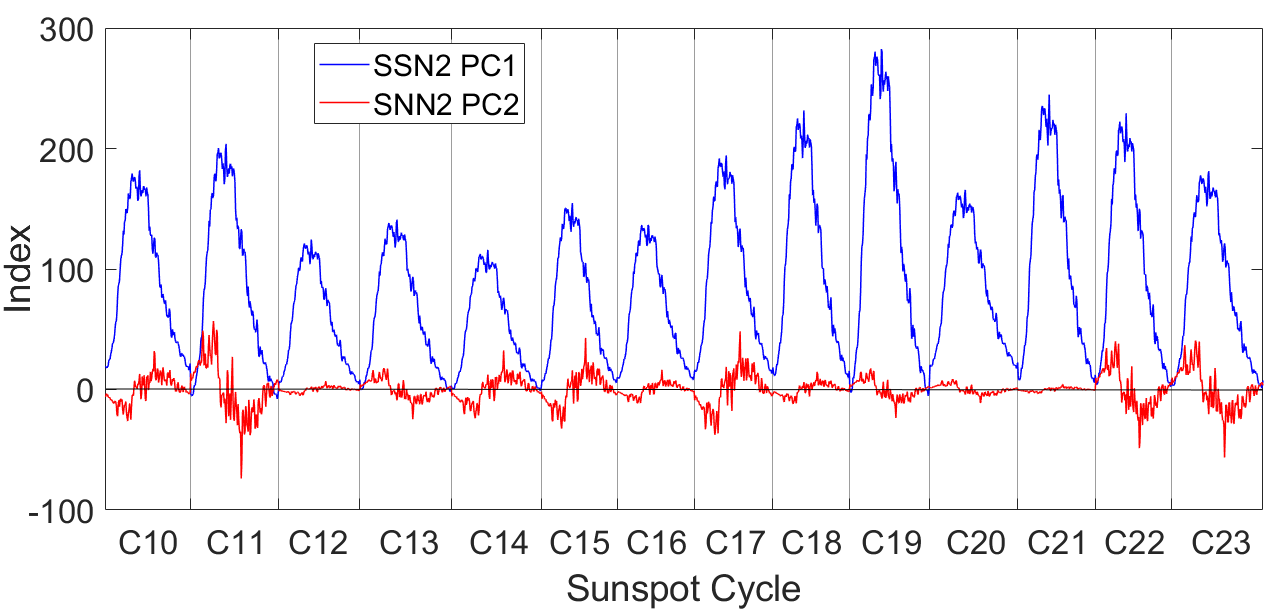}
		\caption{The PC1 (blue) and PC2 (red) time series of SSN for the Solar Cycles 10\,--\,23.}
		\label{fig:R_PC_time_series}
\end{figure}

\begin{figure}
	\centering
	\includegraphics[width=1.0\textwidth]{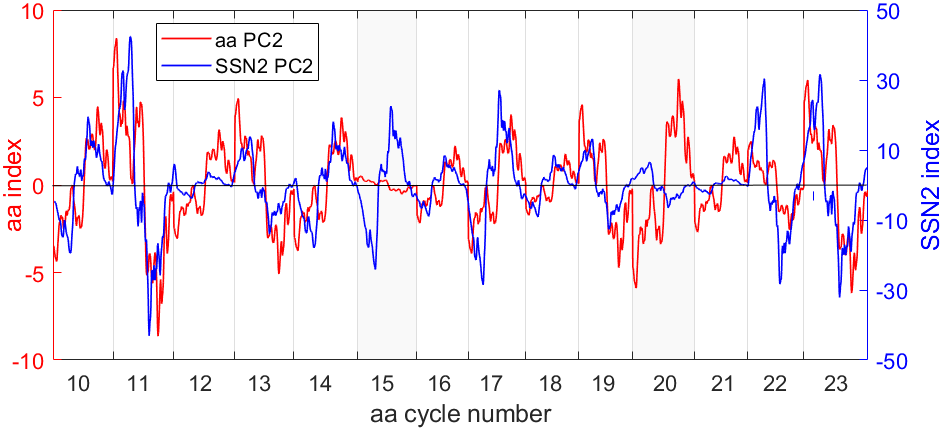}
		\caption{The PC2 time series of the aa-index (red) and SSN (blue) for the Solar Cycles 10\,--\,23.}
		\label{fig:RPC2ts_and_aaPC2ts}
\end{figure}

In the next we study the Solar Cycles 10\,--\,23, because we compare the principal components of sunspot numbers with the principal components of aa-index, which exists only since 1844. In Figure \ref{fig:R_PCs} we show the first two main principal components (PC1 and PC2) of the SSN for the SC10\,--\,SC23, and the corresponding time series PC1 and PC2 proxy in Figure \ref{fig:R_PC_time_series}. In this study we are not so interested in SSN PC1 time series proxy. It is clear that the PC1s of aa-index and SSN are related with each other. More interesting is the relation of the PC2 time series of the aa- and SSN-indices. The relevance of the PC2 is to correct the shape of the cycle when the corresponding cycle differs from the average cycle shape. The main effect of the PC2 is to reduce (positive phase for PC2 in the first half of the cycle) or enhance (negative phase for PC2 in the first half) the activity level of the declining phase with respect to the ascending phase of the cycle \citep{Takalo_2018}. Although PC2 of SSN explains only 3.3\,\% of the variance of the data its significance is more important for some cycles. Figure \ref{fig:RPC2ts_and_aaPC2ts} shows the aa and SSN PC2 time series (seven month smoothed) together. (Note that the vertical lines are minima of the aa-index not SSN-index). The essential feature here is the mutual phase of the aa and SSN PC2 time series. We notice that the phase are same for all other cycles than SC15 and SC20. From these cycles the aa PC2 is rather small for SC15, and SSN PC2 quite small for SC20. Furthermore, cycle 20 is known to be anomalous compared to other aa cycles. The correlation coefficient (CC) between the aa and SSN PC2 time series is best when SSN PC2 is lagged about two years (CC = 0.587 with p$\,<\,10^{-100}$).

\section{Discussion}

The coherence in phase in Figure \ref{fig:RPC2ts_and_aaPC2ts} shows that some part of the shape of the aa cycle PC2 is due to the sunspot number PC2 component. However, most of the aa PC2-component must be caused by other recurrent phenomena in the Sun, and consequently in the solar-wind. We calculated that the average of the second half (descending phase) of the PC2-component of the even cycles is 45.1\,\% compared to the total average of second half for aa-index, but only 13.8\,\% for SSN-index. We thus approximate that about 30\,\% of the shape of aa PC2 is due to sunspot number and the rest for other recurrent processes..  
The first maximum of aa, during sunspot maximum, has been shown to be related to coronal mass ejections (CME), while the second maximum, in the declining phase of SC, is probably related to high-speed streams (HSS) \citep{Gosling_1977, Simon_1986, Simon_1989, Cliver_1996, Echer_2004}. For example, according to Richardson et al. \cite{Richardson_2000} the CME level was unusually low during 1972, i.e, near solar cycle 20 maximum. On the other hand, Gosling et al. \cite{Gosling_1977} found that HSS and streams with velocity in excess of 700 km/sec were common in 1973\,--\,75, i.e. during the declining phase of SC20. These facts are probably the cause of the anomalous shape the cycle 20 aa-index during the maximum and descending phase of the corresponding solar cycle. A counterexample is the strong CME on May 1921 during the descending phase of the Solar Cycle 15, which caused very high values in the magnetometers on Earth \citep{Kappenmann_2006, Hapgood_2019}. This event was, however, so short that it does not appear in the PC2 of the odd-cycle 15. Note, however, that aa cycle 15 has the largest weight of aa PC1-component(see Figure \ref{fig:aa_EOFs}).

\begin{figure}
	\centering
	\includegraphics[width=0.95\textwidth]{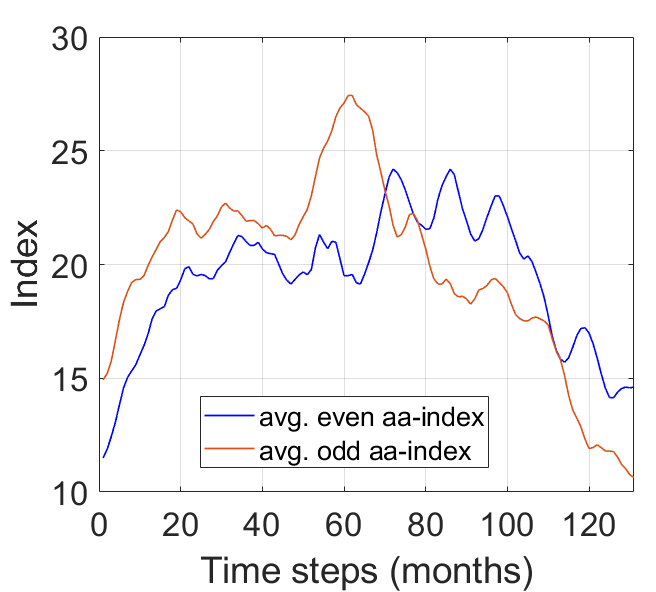}
		\caption{The average shapes of the even (blue) and odd (red) aa cycles C10\,--\,C23.}
		\label{fig:Even-odd_aa_avgs}
\end{figure}

Figure \ref{fig:Even-odd_aa_avgs} shows the average shapes of the even and odd aa cycles during C10\,--\,C23. The even cycles have maxima between 70\,--\,95 months after the start of the aa cycle (related to HSS), while the odd cycles have a single maximum at about 60 months after the start of the cycle (related to CMEs). Note also that the first half of an average even cycle is lower than the first half of an average odd-cycle. \cite{Takalo_2021} has shown that similar average profiles exist in the even and odd cycles of Ap-index.

\section{Conclusion}

We have decomposed monthly aa-index of the cycles 10\,--\,23 using principal component analysis (PCA). Because of the noisiness of the data, we use seven month trapezoidal smoothed aa-index. We show that the first component (PC1) is related to solar sunspot-cycle, and accounts for 41.5\,\%  of the variance of the data. The second component (PC2) is related to 22-year Hale cycle, and explains 23.6\,\% of the variance of the data. The PC3 and higher PCs show shorter periods due to only some individual cycles. PC3 explains, however, still 10.6\,\%, but higher components under 6\,\% of the variance.
The PC1 time series of the cycles 10\,--\,23 aa-index has only one peak in its power spectrum at the period 10.95 years, which is the average solar cycle period for the interval SC10\,--\,SC23. The PC2 time series of the same cycles has a clear peak at period 21.90 (Hale cycle) and a smaller peak at 3/4 of that period. If we remove the cycle pairs 16\,--\,17 and 21\,--\,22 in order to get time series with all succeeding cycles having opposite phases, the 3/4 Hale period disappears, and we get clearer peak at the period 21.90 (note that in this case the last cycle-pair is 20\,--\,23). We also show that the first aa cycle 24 of the ongoing Hale cycle has at least started with a common even cycle phase, i.e. negative phase in the first half of the cycle.
We have also studied the principal component of solar sunspot numbers (SSN) for the Solar Cycles 10\,--\,23, and compared mutual behavior of the PC2 components of aa-index and SSN PCA analyses. We noted that they are in the same phase in all other cycles than Solar Cycle 15 and 20. This shows that at least some part of the different shape of the even and odd aa cycles is due to the solar sunspot cycle. The aa cycle 20 also differs from the other even aa cycles in its shape, especially in anomalously high peaks during its descending phase. It has also, by far, the smallest weight on aa PC1. On the other hand aa cycle 15 has largest weight on the aa PC1 component. These reasons may be the cause of the phase difference of aa cycles 15 and 20 compared to Solar Cycles 15 and 20, respectively.
Even though there is a clear coherence in the phases of the PC2 time series of aa-index and sunspot number index, we estimate that sunspot number explains only 30\,\% the shape of PC2, and the rest is caused by other recurrent events in the Sun/solar-wind. (Note that the PC2 of sunspot numbers accounts only 3.3\,\% of the variance of the corresponding data). The first maximum of aa (typical to odd cycles), during sunspot maximum, has been shown to be related to coronal mass ejections (CME), while the second maximum (typical to even cycles) in the declining phase of SC, is probably related to high-speed streams (HSS) \citep{Gosling_1977, Simon_1986, Simon_1989, Cliver_1996, Echer_2004}. The HSSs raise the activity level such that the minimum between even-odd cycle-pair is always higher than the minimum between succeeding odd-even cycle pair. This can be seen also in the ACF of the detrended aa-index such that Hale cycle related maximum is higher than solar cycle related maximum in its ACF.

\begin{acknowledgments}
The aa-index 1968\,--\,2010 is downloaded from ftp.ngdc.\newline noaa.gov/STP/GEOMAGNETIC\_DATA/AASTAR/, and the aa-index for cycle C24 from ISGI (isgi.unistra.fr). We are grateful to Finnish Meteorological Institute/Nevanlinna for the extended aa-data from 1844 onwards. The dates of cycle minima were obtained from from the National Geophysical Data Center, Boulder, Colorado, USA (ftp.ngdc.noaa.gov). The daily sunspot number data set was downloaded from the SILSO, World Data Center-Sunspot Number and Long-term Solar Observations, Royal Observatory of Belgium, online Sunspot catalog (http://www.sidc.be/silso/datafiles).
\end{acknowledgments}

\flushleft
\textbf{Disclosure of Potential Conflicts of Interest} \newline
\footnotesize {The author declares that there are no conflicts of interest.}

\newpage

\bibliographystyle{spr-mp-sola}
\bibliography{references_JT_aa}  


\end{document}